\documentclass[%
reprint,
superscriptaddress,
 amsmath,amssymb,
 aps,prl
]{revtex4-2}

\usepackage{xcolor}
\usepackage{graphicx}
\usepackage{dcolumn}
\usepackage{bm}

\begin{document}


\title{Casimir nanoparticle levitation in vacuum with broadband perfect magnetic conductor metamaterials}%

\author{Adri\'an E. Rubio L\'opez}
\email{adrianrubiolopez0102@gmail.com}
\affiliation{Birck Nanotechnology Center, School of Electrical and Computer Engineering,\\
 Purdue University, West Lafayette, IN 47907, USA}

\author{Vincenzo Giannini}
\affiliation{Technology Innovation Institute, P.O. Box 9639, Building B04C, Masdar City, Abu Dhabi, United Arab Emirates}
\affiliation{Instituto de Estructura de la Materia (IEM-CSIC), Consejo Superior de Investigaciones Cient\'{i}ficas, Serrano 121, 28006 Madrid, Spain}
\affiliation{Centre of Excellence ENSEMBLE3 Sp.zo.o., Wolczynska Str. 133, 01-919, Warsaw, Poland}
\homepage{http://www.GianniniLab.com}

\date{\today}

\begin{abstract}
%
The levitation of nanoparticles is essential in various branches of research. Casimir forces are natural candidates to tackle it but the lack of broadband metamaterials precluded repulsive forces in vacuum. We show sub-micron nanoparticle levitation in vacuum only based on the design of a broadband metamaterial perfect magnetic conductor surface, where the Casimir force is mostly given by the (quantum) zero-point contribution and compensates the nanoparticle's weight. In the harmonic regime, the volume-independent characteristic frequency depends linearly on Planck's constant $\hbar$.
\end{abstract}

\maketitle

Levitation is an intriguing physical phenomenon that could majorly impact our daily life; a typical example is magnetically levitated trains. Currently, different approaches for levitating objects of different shapes, sizes and materials, and also under a broad variety of scenarios were investigated \cite{Novotny2014,GeraciFridge,Manjavacas2017,BennettBuhmannExp2018,MooreGeraci,Stickler2020,Paulo2020,Quidant2021}. Some approaches exploit the repulsive electric forces perceived by charges of the same sign, while others are based on employing optical potentials or tweezers. Because of its high controllability and hybrid properties, levitated nanoparticles in highly isolated scenarios are objects of high interest since its impact in both technological applications and fundamental science. 
Given their rich phenomenology, nanoparticles are sensitive to fluctuation phenomena, such as Casimir 
forces. The latter were broadly studied as a possible advantageous levitation mechanism~\cite{Kenneth2002, Henkel2005, Levin2010, Milton2012, Sinha2018, Jiang2019, Marchetta2021}, highlighting particularly the pioneer work of Ref.\cite{Boyer1974} in connection to the present work. But strong limitations were found on narrow bandwidths of the materials involved (either for ordinary materials or metamaterials)~\cite{Iannuzzi2003, Kenneth2006, Rosa2008, Rosa2009, Rahi2010}, or the necessity of liquid immersion of the interacting bodies \cite{Rodriguez2010,Sol2015,Rongkuo2019}. Nevertheless, a successful realization in vacuum may lead to a new generation of experiments and applications characterized by minimalistic setups and extreme nanoparticle’s isolation
.

In this Letter we demonstrate the levitation of nanoparticles in vacuum at arbitrary temperature on a sub-micron distance by simply exploiting Casimir interactions with a broadband perfect magnetic conductor (PMC) metamaterial plane surface. We also suggest a possible way to realize such metamaterial.

Taking advantage from the natural repulsive interaction between point objects 
and a PMC surface, we show that even when the PMC property is restricted to an enough-broad 
bandwidth, 
excluding high and low frequencies, the force remains repulsive while its magnitude presents modest variations with respect to the full-bandwidth PMC surface. By opposing this force to the nanoparticle's weight, we show 
sub-micron 
stable levitation for different nanoparticle's materials. For small nanoparticles, 
the levitation dynamics is volume-independent. 
The Casimir 
force is mostly given by the (quantum) 
zero-temperature contribution
, so the levitation mechanism results robust to thermal effects. The resulting asymmetric potential gives anharmonic motion for energies well above the potential's minima. Harmonic dynamics are obtained for low-energies (close to the minima), characterized by a volume-independent frequency with linear dependence on Planck's constant $\hbar$, showing the quantum nature of the phenomenon. 

\begin{figure*}
\includegraphics[width=\linewidth]{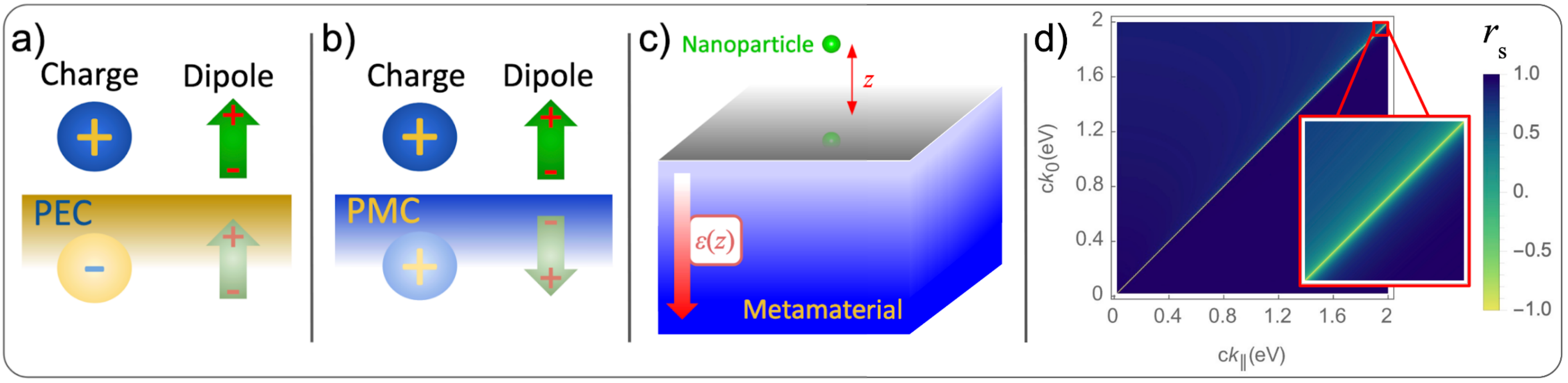}
\caption{\label{fig:0}a,b) Schemes showing the physical intuition for the interaction of basic objects with a PEC (PMC) surface. For the PEC (PMC) case, the charge has an image-charge with the opposite (equal) sign. The force is attractive (repulsive) for the PEC (PMC) case. c) Scheme of the scenario, a quasi-PMC metamaterial surface with a $z-$dependent dielectric constant according to Eq.(\ref{e_profile}), i.e. a gradient index material. At a distance $z$ above the surface, the nanoparticle of radius $R$ is located. d) Reflection coefficients as a function of $\{ck_{0},ck_{\parallel}\}$ for the quasi-PMC with permittivity according to Eq.(\ref{sol_profile}). It is observed that for most of the values $r_{\rm s}\simeq1$ (PMC behavior), while $r_{\rm s}=-1$ (PEC behavior) is obtained in a small region around the light-cone.
}
\end{figure*}

We consider a spherical nanoparticle of radius $R
$ and mass $m=\rho V$, being $\rho$ the density and $V=4\pi R^{3}/3$ the volume. Nanoparticles are well described as electric point-dipoles of polarizability $\alpha(\omega)=V\xi(\omega)$ (Clausius-Mossotti formula), with $\xi(\omega)=3[\varepsilon(\omega)-1]/[\varepsilon(\omega)+2]$, and $\varepsilon(\omega)$ the nanoparticle's material permittivity. The nanoparticle is placed at a distance $z$ from a plane surface with $\hat{n}$ the normal direction 
(see inset Fig.\ref{fig:0}a). 

The physical intuition about the interaction with a PMC plane comes from basic objects in the static case. Schemes with image charges are shown in Figs.\ref{fig:0}b and c. While for a PEC the boundary conditions are $\hat{n}\times\mathbf{E}=0$, for the PMC we have $\hat{n}\times\mathbf{B}=0$. This implies (for any incident angle and frequency) that for a PEC (PMC) we have $r_s=-r_p=\mp 1$. A striking consequence is that while a positive charge interacts with a PEC surface with a negative mirror-charge, for the PMC the mirror-charge is positive; so the force between the PEC (PMC) and a charge is attractive (repulsive). The same intuition is valid for electric dipoles, giving an insight on nanoparticles, although the latter are fluctuating objects. For nanoparticle levitation the spectral broadness of the PMC property is the key-point. Previous works have proved that magnetic resonant metamaterials are not enough~\cite{Iannuzzi2003, Rosa2008, Rosa2009}. The conclusion was that the magnetic properties of resonant metamaterials do not have enough spectral broadness, precluding a concrete realization. The metamaterials inherit their component's resonant nature, implying that the magnetic behavior was limited to a narrow frequency region not enough for levitation effects. Another limiting factor they found is the losses in the metamaterial~\cite{Rosa2008}.

Here we take another approach. We design a long wavelength metamaterial, i.e., a material with the desired properties when the photon wavelength $\lambda$ is much larger than the characteristic spatial scales of the metamaterials (for example, the unit cell in a periodic system). This way, we avoid having a functional material working only on a narrow frequency band. These ideas have been recently applied to obtain metallic transparent metamaterials~\cite{Palmer, Xiao2022}. Thus, we can obtain a quasi-PMC, i.e., a metamaterial behaving as a PMC in a broad range of frequencies.

First, we suggest two possible ways to fabricate such a system. Second, we elucidate the broadband properties of a metamaterial to behave as a PCM. 

Our main objectives are to show that a broadband PMC is possible, demonstrating none fundamental restrictions against it; and also that is enough for realizing nanoparticle levitation.

We aim to obtain a surface with $r_s\approx +1$ in a broad range of frequencies and $k-$vectors. The first idea relies on subwavelength gradient materials~\cite{Shvartsburg2007,Shvartsburg2013} and the existent exact analytical solutions for some refractive index profiles describing light scattering~\cite{Shvartsburg2007, Wait2013}. Such materials have been studied for a long time~\cite{Shvartsburg2007}. In addition, thanks to the continuously improvement of nanofabrication techniques~\cite{Blanco2020}, the nanoscale design of these materials opens a high-interest alternative. A more traditional way to design profiles includes controlling regimes of doping, molecular beam epitaxy, nanoscale porosity variations, fabrication of graded metal-dielectric composites, physical vapor deposition of multiple materials, ion implantation etching, and photolithography.

The first non-trivial profile with an analytical solution was found by Rayleigh a long time ago, in the 1880~\cite{Rayleigh1880}. After that, many other works have contributed to this topic.
Here we are interested in an inverse quadratic variation of the refractive index that leads to Bessel functions to solve the scattering problem~\cite{Wait2013}. Let us assume the following profile for the dielectric constant for the positive semispace $z>0$:
\begin{equation}
    \varepsilon(z) = \varepsilon_1 - \frac{b^2}{(z+L)^2},
    \label{e_profile}
\end{equation}
where $\varepsilon_1, b$ and $L$ are positive constant in our case. Such an inverse square profile has a transition length given by $b$ and $L$. The negative semispace, $z<0$, is assumed to be the vacuum (see figure 1c). The exact solution at such problem for $s$-polarization is given by~\cite{Wait2013}:
\begin{equation}
    r_s = \frac{k_{\perp 0} - s}{k_{\perp 0} + s},
    \label{sol_profile}
\end{equation}
where $k_{\perp 0}=\sqrt{k_0^2-k_\parallel^2}$ is the component of the $k$-vector in free space perpendicular to the plane, $k_\parallel$ is the parallel component (conserved) of the $k$-vector and $k_0=\omega/c$ while $s$ is given by: 
\begin{equation}
    s = \left[ \frac{H^{(2)\prime}_{\nu}(\beta L
    )}{H^{(2)}_\nu (\beta L
    ) } + \frac{1}{2\beta L
    } \right]ik_0\beta,
    \label{s_val}
\end{equation}
where $H^{(2)}_\nu$ and $H^{(2)\prime}_\nu$ are the Hankel function of the second kind and its derivative, $\beta=\sqrt{\varepsilon_1 k_0^2-k_\parallel^2}$ and $\nu=\sqrt{k_0^2 b^2+\tfrac{1}{4}}$. 

By inspection of Eq.~\eqref{sol_profile} and Eq.~\eqref{s_val} we can see that if we have $\varepsilon_1\gg 1$ and $b^2\approx \varepsilon_1 L
^2\gg 1$ this means that $\nu\approx \beta L
\gg1$. In this regime, the ratio of Hankel functions and $1/(2\beta L
)$ go fast to zero, giving $r_s\approx +1$. In order to mimic a PMC, we need to go slowly from lower to high permittivity. 
It is not complex to find high permittivity materials; for example, with self-assembled metal nanoparticles, we can easily get $\varepsilon\sim 100$~\cite{Palmer,Xiao2022} or with a composite material value around $\varepsilon\sim 10^5$ are possible~\cite{Pecharrom2001,Wang2019}. 

For example, choosing $\varepsilon_1 = 100, b=10^3$~nm, and $L= 120$~nm, from Eqs.\eqref{sol_profile} and \eqref{s_val} we obtain that $r_s\approx +1$ in a broad range of frequencies and $k$-vectors, as shown in Fig.\ref{fig:0}d. It turns out that only for a irrelevant sharp region near the light cone $(k_{\perp 0}=0)$ we have $r_s = -1$.

Another possible solution to mimic a PMC could be found in future advances in magnetic nanomaterial composites~\cite{Huang2016,Mourdikoudis2021}. Such materials present strong magnetic effects up to the far infrared but with promising extensions to the near-IR. This can be easily seen from the reflection coefficient between two materials (vacuum/magnetic composite):
\begin{equation}
    r_s = \frac{\mu_1 k_{\perp 0} - k_{\perp 1}}{\mu_1  k_{\perp 0} + k_{\perp 1}}.
    \label{rs}
\end{equation}
Having large values of the magnetic permeability of the nanocomposite ($\mu_1\gg 1$) implies $r_s\approx +1$. 

We want to highlight that there are probably other possible solutions for a broadband PMC but no fundamental reason against it. We hope that more researchers will explore this phenomenon. 
Now we show that nanoparticle levitation is possible with a PMC behavior over a broad region of frequencies and $k$-vectors, while full-spectrum is not necessary.

The force on a nanoparticle arises from the interaction between the surrounding EM field, the plane surface and the nanoparticle considered as a point dipole. A detailed derivation is shown in Sect.I of the Suppl. Mat. Following Refs.\cite{Henkel,AntezzaPRL}, the force over the dipole to the lowest order is $\mathbf{F}(\mathbf{r})\approx\langle\hat{d}_{i}^{(\rm ind)}(t)\nabla\hat{E}_{i}^{(\rm fl)}(\mathbf{r},t)\rangle+\langle \hat{d}_{i}^{(\rm fl)}(t)\nabla\hat{E}_{i}^{(\rm ind)}(\mathbf{r},t)\rangle$, 
where $\mathbf{r}$ is the nanoparticle's position (summation over subscripts is implicit). The first term describes the fluctuations of the field that correlate with the corresponding induced dipole, while the second involves dipole fluctuations and the field they induce. 
In principle, each entity have its own temperature, $\{T_{\rm EM},T_{\rm S},T_{\rm NP}\}$. The force 
over the nanoparticle at a distance $z$ from the surface for a general scenario results:
\begin{equation}
    F_{z}(z)=F_{0}(z)+F_{\rm R}\left(z,T_{\rm EM},T_{\rm NP}\right)+F_{\rm T}\left(z,T_{\rm EM},T_{\rm S}\right),
    \label{FzGeneral}
\end{equation}
where $F_{0}$ stands for the contribution of the zero-point fluctuations, depending on the surface's reflection coefficients $\{r_{s,p}\}$; $F_{\rm R}$ stands for the contribution associated to the surrounding EM field and the nanoparticle radiation also depending on $\{r_{s,p}\}$, while $F_{\rm T}$ relates to the surface's radiation and depends exclusively on its transmission coefficients $\{t_{s,p}\}$.
A metamaterial surface may present frequency cutoffs, having restricted the values of $(\omega,k_{\parallel})$ where $r_{\rm s}= +1$ and $r_{\rm p}= -1$. 
A full-bandwidth PEC (PMC) surface has $r_{\rm s}=\mp 1$, $r_{\rm p}=\pm 1$, while $t_{\rm s,p}= 0$ for every $(\omega,k_{\parallel})$.  
The latter implies that $F_{\rm T}\rightarrow 0$ regardless on the temperatures. In agreement to the intuitive picture of Fig.\ref{fig:0}, in Sect.II of the Suppl. Mat. we show the striking feature $F_{z}^{\rm (PMC)}=-F_{z}^{\rm (PEC)}$. In principle, this theoretically guarantees the levitation of a nanoparticle provided the full-bandwidth PMC property is effective. However, in general metamaterial will present PMC properties on finite bandwidth. We now analyze its impact on the Casimir-Polder force.

\begin{figure}
\includegraphics[width=\columnwidth]{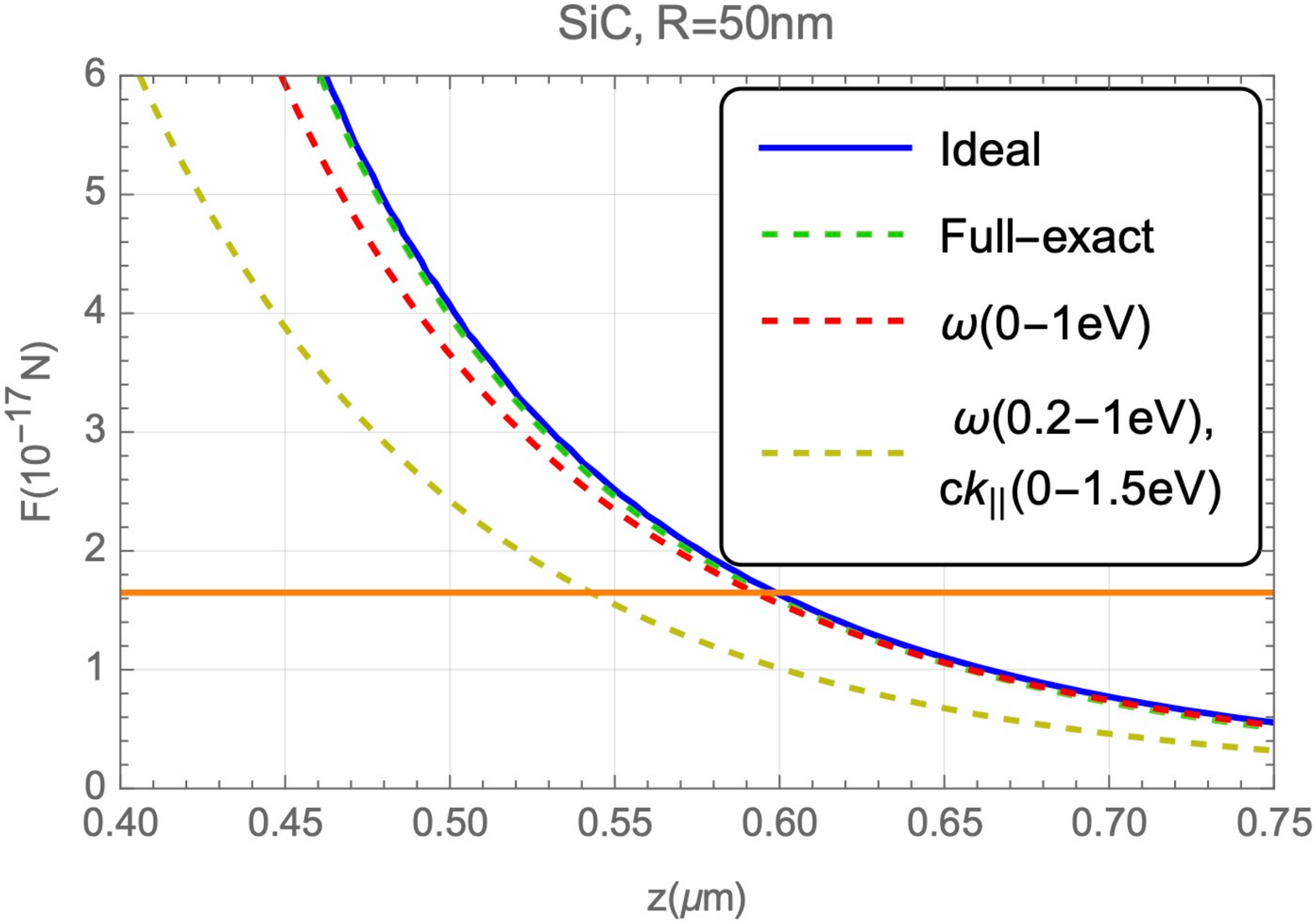}
\caption{\label{fig:ForceCutoff} Casimir force on a SiC nanoparticle ($R=50$~nm) for different upper and lower cutoffs in freqency and/or $k$-vector. Analytical solution to the full-bandwidth ideal PMC (blue line), compared with its numerical solutions (dashed-green line), finite $\omega$-bandwidth (dashed-red line), and finite $\omega-k$-bandwidth (dashed-yellow line). The orange solid horizontal line corresponds to the nanoparticle's weight $mg$, and the intersections with the curves give the levitation position for each case.
}
\end{figure}

For a nanoparticle of $R=50$nm, the weight $mg\sim~10^{-17}$N for common materials such as SiC, Au and Si. The levitation takes place where the Casimir force compensate the weight, as we show below, this occurs for $z<1\mu$m. In the short-distance regime, for which $k_{\rm B}T_{\rm Min}z/[\hbar c]\ll 1$ (with $T_{\rm Min}={\rm min}[T_{\rm Env},T_{\rm NP}]$), the Casimir force is given by the zero-temperature (fully quantum) contribution [see Eq.(S.43) of the Suppl. Mat.]:
\begin{equation}
    F_{z}(z)\approx F_{0}(z)
    .
    \label{FzZeroT}
\end{equation}
This implies that the conclusions obtained from now are robust to thermal effects and relies on the (quantum) zero-point fluctuations (see Sect.IIIA of the Suppl. Mat.). For the full-bandwidth PMC, this contribution reads:
\begin{equation}
    F_{0}(z)\rightarrow F_{0}^{\rm (PMC)}(z)= 3\hbar V I_{0}(z)/(8\pi z^{4}),
    \label{FZeroPoint}
\end{equation}
having $I_{0}(z)\equiv\int_{0}^{+\infty}\frac{d\omega}{2\pi}\xi(i\omega)A(i\omega,z)e^{-2\frac{\omega}{c}z}$, with $A(i\omega,z)=\sum_{n=0}^{3}\frac{1}{n!}\left(2\frac{\omega}{c}z\right)^{n}$. In Fig.~\ref{fig:ForceCutoff} we show the Casimir force acting on a SiC nanoparticle of $R=50$nm for different upper and lower frequency/$k$-vectors cutoffs combinations numerically obtained by employing Eq.(\ref{sol_profile}) for the surface (dashed-red and dashed-yellow curves), as well as the exact analytical and numerical cases for case the full-bandwidth PMC given by Eq.(\ref{FZeroPoint}) (blue solid and dashed-green curves).
For SiC we employed a permittivity model $\varepsilon_{\rm SiC}(\omega)=\varepsilon_{\infty}(\omega_{\rm L}^{2}-\omega^{2}-i\gamma\omega)/(\omega_{\rm T}^{2}-\omega^{2}-i\gamma\omega)$, where $\omega_{\rm L} =18.253~10^{13}{\rm s}^{-1}$, $\omega_{\rm T} =14.937~10^{13}{\rm s}^{-1}$, $\gamma=8.966~10^{11}{\rm s}^{-1}$ and $\varepsilon_{\infty}=6.7$. According to the basic physical intuition conveyed in Fig. \ref{fig:0}, a repulsive force acts on the nanoparticle  even in the broadband PMC case, i.e. the not full-bandwidth case (dashed red and yellow lines). A broader interval in frequencies increases the repulsion at every distance from the surface. The maximum repulsion is achieved for the full-bandwidth PCM (blue solid and dash green lines). Furthermore, in Sect.III of the Suppl. Mat., we show that for a SiC nanoparticle the force by a full-bandwidth ideal PMC surface, $F_{0}^{\rm (PMC)}$ (blue solid curve), can be approximated by:
\begin{equation}
    F_{0}^{\rm (PMC)}(z)\approx \frac{9\hbar cV}{8\pi^{2}z^{5}}\frac{(\varepsilon_{\infty}-1)}{(\varepsilon_{\infty}+2)}.
    \label{FZeroFullPMC}
\end{equation}
When including cutoffs, the differences with respect to the full-bandwidth case are not critical. This allow us to take the full-bandwidth PMC case as representative for our analysis. In other words, our metamaterial with a finite bandwidth is mostly approximated by of a full-bandwidth PMC. 

From now on, we focus on the latter for continuing the analysis, while considering that a finite bandwidth effectively decreases the force in a moderate magnitude. A striking feature is that in all the cases the nanoparticle's weight (orange solid line, taking $\rho=3210{\rm Kg}/{\rm m}^{3}$) is compensated by the Casimir-Polder force in distances between $\sim 0.53-0.6\mu$m. All these results can be extended to Au and Si nanoparticles. In Sect.IIIB of the Suppl. Mat. we show that levitation for SiC, Au and Si nanoparticles are possible between the 0.4-0.7$\mu$m for the full-bandwidth PMC.

\begin{figure}
\includegraphics[width=\columnwidth]{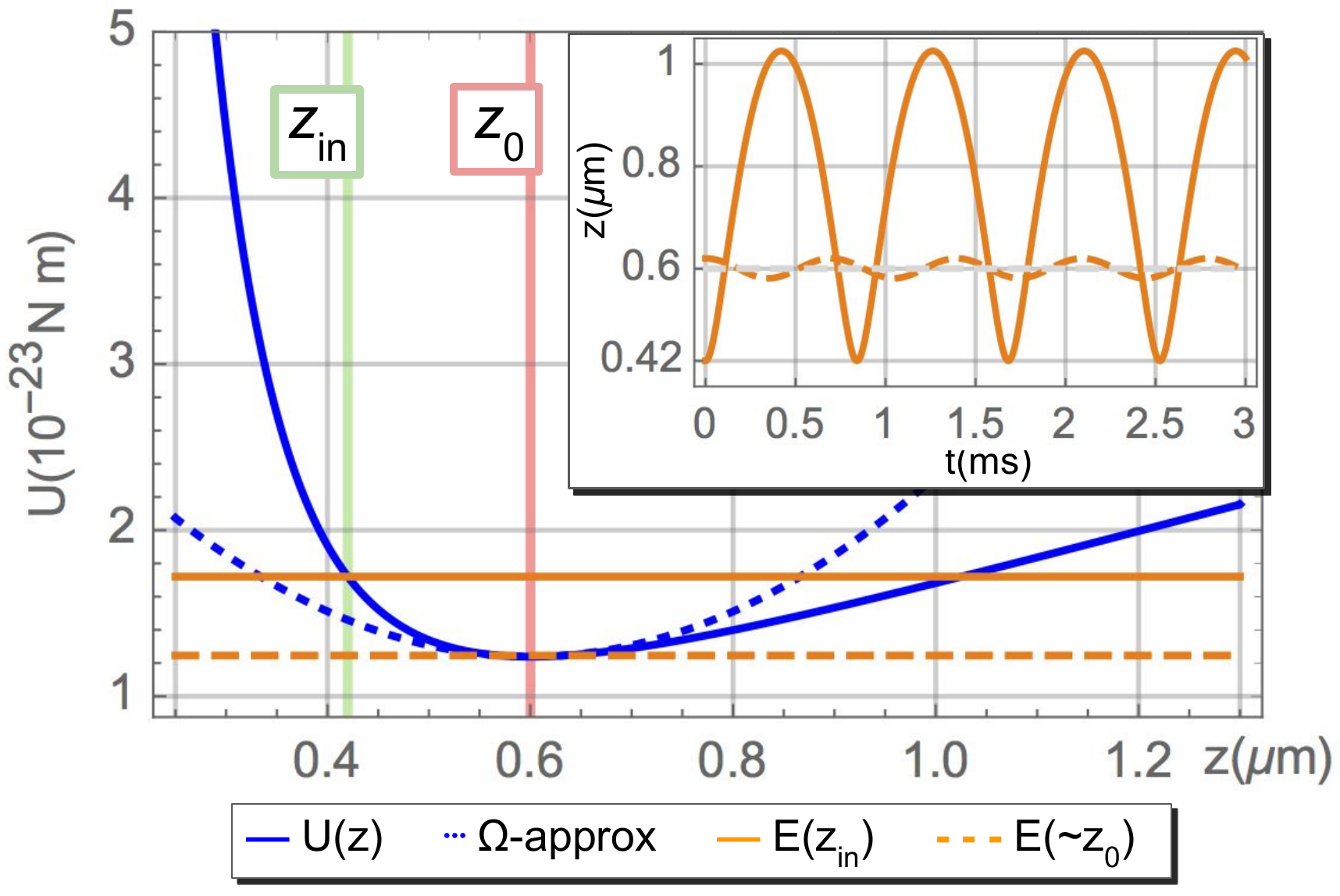}
\caption{\label{fig:LevitationTriPanel} SiC nanoparticle ($R=50$nm) levitation, with an equilibrium position $z_{0}\approx0.6\mu$m. The blue solid curve corresponds to the total potential $U$ on the nanoparticle. The dotted blue curve is its harmonic approximation around $z_{0}$ (vertical red line). The orange solid (dashed) line corresponds to an energy level defined by $z(t_{\rm in})=z_{\rm in}$ [$z(t_{\rm in})\approx z_{0}$] and $\dot{z}(t_{\rm in})=0$. Inset: Trajectories of the nanoparticle $z(t)$ for both energy levels, giving anharmonic and harmonic motions, respectively.}
\end{figure}

By opposing the Casimir force and the nanoparticle's weight, the mechanical equilibrium position $z_{0}$ is given by $F_{z}(z_{0})=mg$. As it was shown, we have $R\ll z_{0}\approx0.6\mu$m irrespective of the temperatures, and in agreement to the point-dipole approximation. Furthermore, as for small nanoparticles the radiation reaction correction entering the force through the polarizability is negligible, the equilibrium position $z_{0}$ results independent of the volume, $V$. Thus, the levitation of nanoparticles interacting with a PMC surface shows to be feasible. The total potential perceived by the nanoparticle is calculated from the combination of the Casimir force with the weight as $U(z)=-\int_{z}^{+\infty}F_{z}(z')dz'+mgz$, which minimum is $\{z_{0},U_{0}\}$. This is shown in Fig.\ref{fig:LevitationTriPanel} as the blue solid curve.
Given the potential shape, levitation within a maximum and minimum distances is ensured. As the potential is not symmetric around $z_{0}$, the motion on the $z$ direction is, in general, anharmonic. As long as the nanoparticle's motion is limited to distances below 1$\mu$m, thermal effects do not take a role. This is the case for an energy level $E(z_{\rm in})=U(z_{\rm in})$, defined by the initial conditions $\{z(t_{\rm in})=z_{\rm in},\dot{z}(t_{\rm in})=0\}$ (orange solid curve corresponding to $z_{\rm in}=0.42\mu$m). The actual motion of the nanoparticle is obtained from the equation of motion $m\ddot{z}(t)=F_{z}(z(t))-mg$. The anharmonic trajectory of the nanoparticle is shown in the inset of Fig.\ref{fig:LevitationTriPanel}, presenting a period $\sim 1$ms. For energy levels $E(\sim z_{0})\approx U_{0}$, the motion is harmonic, as shown by the orange dashed curves (taking $z_{\rm in}=0.57\mu{\rm m}$). The potential verifies $U(z)\approx m\Omega^{2}(z-z_{0})^{2}/2$ (blue dotted curve), while the equation of motion is $\delta\ddot{z}(t)=-\Omega^{2}\delta z$, provided $\delta z\equiv z-z_{0}\ll z_{0}$. The characteristic frequency reads $\Omega\equiv[-(1/m)(dF_{z}/dz)|_{z_{0}}]^{1/2}$. According to Eq.(\ref{FZeroFullPMC}), in the short-distance regime the force depends linearly on $\hbar$ as an evidence of its quantum nature, so the frequency results:
\begin{equation}
\Omega^{2}
\approx\frac{5g}{z_{0}}=\frac{45\hbar c}{8\pi^{2}\rho z_{0}^{6}}\frac{(\varepsilon_{\infty}-1)}{(\varepsilon_{\infty}+2)},
\end{equation}
which, as $z_{0}$, is independent of $V$. For a SiC nanoparticle, $\Omega\approx 9013~{\rm s^{-1}}$ (or equivalently, $\nu=\Omega/[2\pi]\approx 1.4~{\rm KHz}$), giving a period of 0.7ms. Remarkably, while the first expression for the frequency $\Omega$ is minimal, the second one is linear on $\hbar$. 

All in all, in this Letter we have analyzed the possibility of nanoparticle levitation by exploiting the Casimir force in combination with a novel PMC metamaterial surface. We suggested two possible design of a broadband PMC metamaterial based on a gradient materials or magnetic nanocomposites, although other equivalent options could also be explored~\cite{AshwinPaper}. We show that the weight of a material nanoparticle can be compensated by the Casimir force, ensuring levitation.
Having the nanoparticle confined on the short-distances regime $z<1\mu$m, the mechanism is sustained by the (quantum) zero-temperature contribution. Thus, the mechanism shows to be robust to thermal effects and feasible for experiments in vacuum. We finally show that quantum effects on the harmonic dynamics (around the equilibrium point $z_{0}$) gives a characteristic frequency with a linear dependence on Planck's constant $\hbar$ and independent of the nanoparticle's volume. Our work opens the way on the experimental and technological sides for nanoparticles levitation in vacuum, and also for achieving levitation of larger objects as parallel plates in frictionless contact.

\begin{acknowledgments}
V.G. thanks the “ENSEMBLE3 - Centre of Excellence for nanophotonics, advanced materials and novel crystal growth-based technologies” project (GA No. MAB/2020/14) and the European Union’s Horizon 2020 research and innovation programme Teaming for Excellence (GA. No. 857543).
\end{acknowledgments}

\bibliography{References}




\end{document}


\title{Casimir nanoparticle levitation in vacuum with broadband perfect magnetic conductor metamaterials}%

\author{Adri\'an E. Rubio L\'opez}
\email{adrianrubiolopez0102@gmail.com}
\affiliation{Birck Nanotechnology Center, School of Electrical and Computer Engineering,\\
 Purdue University, West Lafayette, IN 47907, USA}



\author{Vincenzo Giannini}
\affiliation{Technology Innovation Institute, P.O. Box 9639, Building B04C, Masdar City, Abu Dhabi, United Arab Emirates}
\affiliation{Instituto de Estructura de la Materia (IEM-CSIC), Consejo Superior de Investigaciones Cient\'{i}ficas, Serrano 121, 28006 Madrid, Spain}
\affiliation{Centre of Excellence ENSEMBLE3 sp. z o.o., Wolczynska 133, Warsaw, 01-919, Poland}
\homepage{http://www.GianniniLab.com}

\maketitle

\section{Casimir-Polder force on a nanoparticle}

Our general scenario of interest consists in a nanoparticle close to a surface in a out of thermal equilibrium situation, meaning that the nanoparticle, the surface and the enviromental EM field have different temperatures ($T_{\rm NP}$, $T_{\rm S}$ and $T_{\rm EM}$ respectively). As the nanoparticle is considered to be of a typical size smaller than the typical wavelength of the surrounding radiation (with energies in the orders of $T_{\rm S,EM}$, considering $k_{\rm B}\equiv 1$), the nanoparticle interacts with the EM field through its dipole moment $\mathbf{d}$.

Therefore, following Ref.\cite{Henkel}, the force over the dipole reads:
\begin{equation}
\mathbf{F}\left(\mathbf{r}\right)=\left\langle\hat{d}_{i}^{(\rm ind)}(t)\nabla\hat{E}_{i}^{(\rm fl)}\left(\mathbf{r},t\right)\right\rangle+\left\langle \hat{d}_{i}^{(\rm fl)}(t)\nabla\hat{E}_{i}^{(\rm ind)}\left(\mathbf{r},t\right)\right\rangle,
\label{ForceDipole}
\end{equation}
\noindent where $\mathbf{r}$ stands for the nanoparticle's position.

The first term describes the fluctuations of the field that correlate with the corresponding induced dipole, while the second involves dipole fluctuations and the field they induce. Crossed terms do not appear since there are no crossed correlations between the dipole and the field, since they originate from different physical systems.

On one hand, the dipole induced by the field fluctuations is given by the particle's polarizability $\alpha(\omega)$:
\begin{equation}
\hat{\mathbf{d}}^{(\rm ind)}(\omega)=\varepsilon_{0}\alpha(\omega)\hat{\mathbf{E}}\left(\omega,\mathbf{r}\right),
\end{equation}
\noindent where $\hat{\mathbf{E}}\left(\omega,\mathbf{r}\right)$ is the total electric field at the dipole's position. Working at lowest order in the polarizability, we can ignore the field scattered by the nanoparticle and replace it with $\hat{\mathbf{E}}^{(\rm fl)}\left(\omega,\mathbf{r}\right)$. Otherwise, we could work with a `dressed' polarizability. 

On the other hand, the field induced by the dipole fluctuations is written in terms of the EM Green tensor:
\begin{equation}
\hat{\mathbf{E}}^{(\rm ind)}\left(\omega,\mathbf{x}\right)=\mathbf{G}\left(\omega,\mathbf{x},\mathbf{r}\right)\cdot\hat{\mathbf{d}}\left(\omega\right),
\end{equation}
\noindent having that $\mathbf{x}$ is an observation point. In analogy as before, to the lowest order, we can identify the total dipole with its fluctuating component $\hat{\mathbf{d}}^{(\rm fl)}$.

Moreover, by considering that:
\begin{equation}
\left\langle\hat{A}(t)\hat{B}(t)\right\rangle=\int\frac{d\omega}{2\pi}\frac{d\omega'}{2\pi}~e^{i(\omega'-\omega)t}\left\langle \hat{A}(\omega)\hat{B}^{\dag}(\omega')\right\rangle,
\end{equation}
\noindent we can proceed to write each term of the force given in Eq.(\ref{ForceDipole}). For these calculations, we require the correlations of the fluctuating dipole and of the different contributions of the field.

For the dipole fluctuations we appeal to the fluctuation-dissipation theorem which, assuming the nanoparticle to be at temperature $T_{\rm NP}$, gives:
\begin{equation}
\left\langle\hat{\mathbf{d}}^{(\rm fl)}(\omega)\left[\hat{\mathbf{d}}^{(\rm fl)}(\omega')\right]^{\dag}\right\rangle=2\pi\delta(\omega-\omega')\delta_{ij}\frac{2\hbar\varepsilon_{0}}{\left(1-e^{-\hbar\omega/(k_{B}T_{\rm NP})}\right)}{\rm Im}\left[\alpha(\omega)\right].
\end{equation}

It is important to notice that this expression applies for positive and negative frequencies. For $\omega>0$, we have $(1-{\rm Exp}[-\hbar\omega/(k_{B}T_{i}
)])^{-1}=1+\overline{n}_{i}(\omega
)$, with $\overline{n}_{i}(\omega)$ the Bose-Einstein occupation number at temperature $T_{i}$. For $\omega<0$, we have $(1-{\rm Exp}[-\hbar\omega/(k_{B}T_{i})])^{-1}=-\overline{n}_{i}(|\omega|)$. The minus sign is compensated by the oddness of the imaginary part of the polarizability. In our notation, we will keep present in the subscripts the value of the temperatures $T_{i}$ at $\overline{n}_{i}$ in order to avoid any confusion since we are dealing with nonequilibrium situations involving multiple temperatures.

Therefore, we can prove that for the dipole fluctuations term we have:
\begin{eqnarray}
\left\langle \hat{d}_{i}^{(\rm fl)}(t)\nabla\hat{E}_{i}^{(\rm ind)}\left(\mathbf{r},t\right)\right\rangle&=&2\hbar\varepsilon_{0}\int_{-\infty}^{+\infty}\frac{d\omega}{2\pi}\frac{{\rm Im}\left[\alpha(\omega)\right]}{\left(1-e^{-\hbar\omega/(k_{B}T_{\rm NP})}\right)}\nabla\left[G_{ii}^{*}\left(\omega,\mathbf{r},\mathbf{r}'\right)\right]\Big|_{\mathbf{r}'=\mathbf{r}}\\
&=&2\hbar\varepsilon_{0}\int_{0}^{+\infty}\frac{d\omega}{2\pi}{\rm Im}\left[\alpha(\omega)\right]\left(\nabla\left[G_{ii}^{*}\left(\omega,\mathbf{r},\mathbf{r}'\right)\right]|_{\mathbf{r}'=\mathbf{r}}+2\overline{n}_{\rm NP}(\omega
){\rm Re}\left[\nabla\left[G_{ii}\left(\omega,\mathbf{r},\mathbf{r}'\right)\right]|_{\mathbf{r}'=\mathbf{r}}\right]\right),\nonumber
\end{eqnarray}
\noindent where we have considered $\coth\left[\hbar\omega/(2k_{B}T_{i})\right]=1+2\overline{n}_{i}(\omega)$ as $\omega>0$.

Furthermore, in our particular case of a half-space, as the source and observation point are equal to the position of the dipole ($\mathbf{r}$) in free space, the EM Green tensor will be constituted by two parts: $G_{ij}(\omega,\mathbf{r},\mathbf{r}')=G_{ij}^{(\rm Free)}(\omega,\mathbf{r},\mathbf{r}')+G_{ij}^{(\rm Refl)}(\omega,\mathbf{r},\mathbf{r}')$, where the first term correspond to the free field Green tensor and the second one to the reflection from the half-space. However, it can be checked that, under the derivative, the free field part vanishes when setting $\mathbf{r}'=\mathbf{r}$ due to isotropy (which means that $G_{ij}^{(\rm Free)}(\omega,\mathbf{r},\mathbf{r}')=G_{ij}^{(\rm Free)}(\omega,\mathbf{r}-\mathbf{r}')$) and thus we have that $\nabla\left[G_{ij}\left(\omega,\mathbf{r},\mathbf{r}'\right)\right]|_{\mathbf{r}'=\mathbf{r}}=\nabla[G_{ij}^{(\rm Refl)}\left(\omega,\mathbf{r},\mathbf{r}'\right)]|_{\mathbf{r}'=\mathbf{r}}$.

On the other hand, for the field fluctuations' contribution, we can write:
\begin{equation}
\left\langle\hat{d}_{i}^{(\rm ind)}(t)\nabla\hat{E}_{i}^{(\rm fl)}\left(\mathbf{r},t\right)\right\rangle=\int\frac{d\omega}{2\pi}\frac{d\omega'}{2\pi}~e^{i(\omega'-\omega)t}\varepsilon_{0}\alpha(\omega)\delta_{ij}\nabla'\left[\left\langle\hat{E}_{i}^{(\rm fl)}\left(\omega,\mathbf{r}\right)\left[\hat{E}_{j}^{(\rm fl)}\left(\omega',\mathbf{r}'\right)\right]^{\dag}\right\rangle\right]\Big|_{\mathbf{r}'=\mathbf{r}}.
\end{equation}

Thus, the correlations of the electric field are required, i.e., the field fluctuations. These have two contributions: one coming from the thermal currents in the material half-space, and a second one coming from the field fluctuations in the empty half-space:
\begin{equation}
\left\langle\hat{E}_{i}^{(\rm fl)}\left(\omega,\mathbf{r}\right)\left[\hat{E}_{j}^{(\rm fl)}\left(\omega',\mathbf{r}'\right)\right]^{\dag}\right\rangle=2\pi\delta\left(\omega-\omega'\right)\left[W_{ij}\left(\omega,T_{\rm S},\mathbf{r},\mathbf{r}'\right)+V_{ij}\left(\omega,T_{\rm EM},\mathbf{r},\mathbf{r}'\right)\right].
\end{equation}

At this point, it is worth noting that each contribution ($W_{ij},V_{ij}$) can be splitted as a sum of a zero-point fluctuation plus a thermal contribution in the following way:
\begin{equation}
W_{ij}\left(\omega,T_{\rm S},\mathbf{r},\mathbf{r}'\right)=\left\{\theta(\omega)\left[1+\overline{n}_{\rm S}(\omega
)\right]-\theta(-\omega)\overline{n}_{\rm S}(|\omega|)\right\}W_{ij}\left(\omega,\mathbf{r},\mathbf{r}'\right),
\end{equation}
\begin{equation}
V_{ij}\left(\omega,T_{\rm EM},\mathbf{r},\mathbf{r}'\right)=\left\{\theta(\omega)\left[1+\overline{n}_{\rm EM}(\omega
)\right]-\theta(-\omega)\overline{n}_{\rm EM}(|\omega|)\right\}V_{ij}\left(\omega,\mathbf{r},\mathbf{r}'\right),
\end{equation}
\noindent where the second terms vanish for the corresponding zero temperature limit ($T_{\rm S},T_{\rm EM}\rightarrow 0$).

At the same time, the first two terms combine to give the zero-temperature fluctuation-dissipation theorem:
\begin{equation}
W_{ij}\left(\omega,0,\mathbf{r},\mathbf{r}'\right)+V_{ij}\left(\omega,0,\mathbf{r},\mathbf{r}'\right)=2\hbar\theta(\omega){\rm Im}\left[G_{ij}\left(\omega,\mathbf{r},\mathbf{r}'\right)\right],
\end{equation}
\noindent where the Heaviside function from the limit:
\begin{equation}
\theta(\omega)=\lim_{T\rightarrow 0}\frac{1}{1-e^{-\hbar\omega/(k_{B}T)}}.
\end{equation}

Equivalently for $W_{ij}\left(\omega,\mathbf{r},\mathbf{r}'\right),V_{ij}\left(\omega,\mathbf{r},\mathbf{r}'\right)$, we can write:
\begin{equation}
W_{ij}\left(\omega,\mathbf{r},\mathbf{r}'\right)+V_{ij}\left(\omega,\mathbf{r},\mathbf{r}'\right)=2\hbar{\rm Im}\left[G_{ij}\left(\omega,\mathbf{r},\mathbf{r}'\right)\right],
\end{equation}

Thus, we have:
\begin{equation}
V_{ij}\left(\omega,T_{\rm EM},\mathbf{r},\mathbf{r}'\right)=\left\{\theta(\omega)\left[1+\overline{n}_{\rm EM}(\omega)\right]-\theta(-\omega)\overline{n}_{\rm EM}(|\omega|)\right\}\left[2\hbar{\rm Im}\left[G_{ij}\left(\omega,\mathbf{r},\mathbf{r}'\right)\right]-W_{ij}\left(\omega,\mathbf{r},\mathbf{r}'\right)\right],
\end{equation}

Now, following Ref.\cite{Henkel}, from Lifshitz theory, considering that the radiation generated by the material half-space is due to fluctuating sources playing the role of a polarization field, and assuming local thermal equilibrium, we use the fluctuation-dissipation theorem to calculate the contribution of the material plane to the field fluctuations, obtaining:
\begin{equation}
W_{ij}\left(\omega,T_{\rm S},\mathbf{r},\mathbf{r}'\right)=2\hbar\varepsilon_{0}\frac{{\rm Im}\left[\varepsilon_{\rm S}(\omega)\right]}{\left(1-e^{-\hbar\omega/(k_{B}T_{\rm S})}\right)}S_{ij}\left(\omega,\mathbf{r},\mathbf{r}'\right),
\end{equation}
\noindent with $\varepsilon_{\rm S}(\omega)$ the permittivity of the material half-space and:
\begin{equation}
S_{ij}\left(\omega,\mathbf{r},\mathbf{r}'\right)=\int_{V_{\rm S}}d\mathbf{x}~G_{ik}\left(\omega,\mathbf{r},\mathbf{x}\right)G_{jk}^{*}\left(\omega,\mathbf{r}',\mathbf{x}\right).
\end{equation}



This allow us to prove the fundamental property of the functions $W_{ij},V_{ij}$:
\begin{equation}
W_{ij}(-\omega,\mathbf{r},\mathbf{r}')=-W_{ij}^{*}(\omega,\mathbf{r},\mathbf{r}')~~~,~~~V_{ij}(-\omega,\mathbf{r},\mathbf{r}')=-V_{ij}^{*}(\omega,\mathbf{r},\mathbf{r}'),
\end{equation}
\noindent provided also that $G_{ij}(-\omega,\mathbf{r},\mathbf{r}')=G_{ij}^{*}(\omega,\mathbf{r},\mathbf{r}')$.

The electric field correlation reads:
%
%
\begin{eqnarray}
\left\langle\hat{E}_{i}^{(\rm fl)}\left(\omega,\mathbf{r}\right)\left[\hat{E}_{j}^{(\rm fl)}\left(\omega',\mathbf{r}'\right)\right]^{\dag}\right\rangle&=&2\pi\delta\left(\omega-\omega'\right)\Bigg(2\hbar\left[\theta(\omega)\left(1+\overline{n}_{\rm EM}(\omega)\right)-\theta(-\omega)\overline{n}_{\rm EM}(|\omega|)\right]{\rm Im}\left[G_{ij}(\omega,\mathbf{r},\mathbf{r}')\right]\\
&&+\left\{\theta(\omega)\left[\overline{n}_{\rm S}(\omega)-\overline{n}_{\rm EM}(\omega)\right]-\theta(-\omega)\left[\overline{n}_{\rm S}(|\omega|)-\overline{n}_{\rm EM}(|\omega|)\right]\right\}W_{ij}(\omega,\mathbf{r},\mathbf{r}')\Bigg).\nonumber
\end{eqnarray}

The contribution to the force then reads:
%
%
\begin{eqnarray}
\left\langle\hat{d}_{i}^{(\rm ind)}(t)\nabla\hat{E}_{i}^{(\rm fl)}\left(\mathbf{r},t\right)\right\rangle&=&
2\varepsilon_{0}\int_{0}^{+\infty}\frac{d\omega}{2\pi}\Bigg[\hbar\left[\alpha(\omega)+2{\rm Re}\left[\alpha(\omega)\right]\overline{n}_{\rm EM}(\omega)\right]{\rm Im}\left[\nabla\left[G_{ii}(\omega,\mathbf{r},\mathbf{r}')\right]|_{\mathbf{r}'=\mathbf{r}}\right]\nonumber\\
&&+\left[\overline{n}_{\rm S}(\omega)-\overline{n}_{\rm EM}(\omega)\right]{\rm Re}\left[\alpha(\omega)\nabla'\left[W_{ii}(\omega,\mathbf{r},\mathbf{r}')\right]|_{\mathbf{r}'=\mathbf{r}}\right]\Bigg],
\end{eqnarray}
\noindent where we have used that $\nabla\left[G_{ii}\left(\omega,\mathbf{r},\mathbf{r}'\right)\right]|_{\mathbf{r}'=\mathbf{r}}=\nabla'\left[G_{ii}\left(\omega,\mathbf{r},\mathbf{r}'\right)\right]|_{\mathbf{r}'=\mathbf{r}}$.

Considering this last equation, we can write the force over the dipole as:
\begin{eqnarray}
&&\mathbf{F}\left(\mathbf{r}\right)=
2\varepsilon_{0}\int_{0}^{+\infty}\frac{d\omega}{2\pi}\Bigg[\hbar\coth\left(\frac{\hbar\omega}{2k_{B}T_{\rm EM}}\right){\rm Re}\left[\alpha(\omega)\right]{\rm Im}\left[\nabla\left[G_{ii}\left(\omega,\mathbf{r},\mathbf{r}'\right)\right]|_{\mathbf{r}'=\mathbf{r}}\right]\\
&&+~\hbar\coth\left(\frac{\hbar\omega}{2k_{B}T_{\rm NP}}\right){\rm Im}\left[\alpha(\omega)\right]{\rm Re}\left[\nabla\left[G_{ii}(\omega,\mathbf{r},\mathbf{r}')\right]|_{\mathbf{r}'=\mathbf{r}}\right]
+\left[\overline{n}_{\rm S}(\omega)-\overline{n}_{\rm EM}(\omega)\right]{\rm Re}\left[\alpha(\omega)\nabla'\left[W_{ii}(\omega,\mathbf{r},\mathbf{r}')\right]|_{\mathbf{r}'=\mathbf{r}}\right]\Bigg],\nonumber
\label{ForceIntegralForm}
\end{eqnarray}
\noindent where the force results to be explicitly real.

Now, we have to evaluate the EM Green tensor and the function $W_{ii}$ for our particular case. Following Ref.\cite{Henkel}, for the case of a half-space with surface at $z=0$, the reflection part of the EM Green tensor is given by:
\begin{equation}
G_{ij}^{(\rm Refl)}(\omega,\mathbf{r},\mathbf{r}')=\int\frac{d^{2}K}{(2\pi)^{2}}~e^{i\mathbf{K}\cdot\left(\mathbf{R}-\mathbf{R}'\right)}~e^{i\gamma(z+z')}~g_{ij}\left(\omega,\mathbf{K}\right),
\end{equation}
\noindent with:
\begin{equation}
g_{ij}\left(\omega,\mathbf{K}\right)=\frac{i\omega^{2}}{2\varepsilon_{0}c^{2}\gamma}\sum_{\mu=s,p}e_{\mu,i}^{(r)}e_{\mu,j}^{(i)}r_{\mu},
\end{equation}
\noindent where $\mathbf{K}=(k_{x},k_{y},0)$ and $\gamma=\sqrt{\omega^{2}/c^{2}-K^{2}}$ is the wavevector component perpendicular to the interface. The unit vectors $\mathbf{e}_{\mu}^{(i,r)}$ describe the polarization of plane waves incident from the vacuum half-space $z>0$ and reflected from the interface, respectively, with reflection coefficient $r_{\mu}$. The polarization vectors are given by $\mathbf{e}_{s}^{(i,r)}=\hat{\mathbf{K}}\times\mathbf{e}_{z}$, $\mathbf{e}_{p}^{(i)}=(K\mathbf{e}_{z}+\gamma\hat{\mathbf{K}})c/\omega$, and $\mathbf{e}_{p}^{(r)}=(K\mathbf{e}_{z}-\gamma\hat{\mathbf{K}})c/\omega$, where $\hat{\mathbf{K}}=\mathbf{K}/K$ is a unit vector.

On the other hand, the function $W_{ij}$ is given by:
\begin{equation}
W_{ij}(\omega,T,\mathbf{r},\mathbf{r}')=\frac{2\hbar\omega^{3}}{3\pi\varepsilon_{0}c^{3}\left(1-e^{-\frac{\hbar\omega}{k_{B}T}}\right)}\int\frac{d^{2}K}{(2\pi)^{2}}~e^{i\mathbf{K}\cdot\left(\mathbf{R}-\mathbf{R}'\right)}~e^{i(\gamma z-\gamma^{*}z')}~\mathbb{W}_{ij}\left(\omega,\mathbf{K}\right)\equiv\frac{1}{\left(1-e^{-\frac{\hbar\omega}{k_{B}T}}\right)}W_{ij}(\omega,\mathbf{r},\mathbf{r}'),
\end{equation}
\noindent with:
\begin{equation}
\mathbb{W}_{ij}\left(\omega,\mathbf{K}\right)=\frac{3\pi c}{4\omega}\frac{{\rm Re}(\gamma_{\rm S})}{|\gamma_{\rm S}|^{2}}\sum_{\mu=s,p}e_{\mu,i}^{(t)}e_{\mu,j}^{(t) *}|\mathbf{e}_{\mu}^{(\rm S)}|^{2}|t_{\mu}|^{2},
\end{equation}
\noindent where $\gamma_{\rm S}=\sqrt{\varepsilon_{\rm S}\omega^{2}/c^{2}-K^{2}}$ is the perpendicular wavevector component inside the substrate. The substrate and vacuum polarization vectors are denoted $\mathbf{e}_{\mu}^{({\rm S},t)}$, respectively, with the transmission coefficient $t_{\mu}$. The polarization vectors are given by $\mathbf{e}_{\mu}^{(t)}=\mathbf{e}_{\mu}^{(r)}$, $\mathbf{e}_{s}^{(\rm S)}=\hat{\mathbf{K}}\times\mathbf{e}_{z}$ and $\mathbf{e}_{p}^{(\rm S)}=(K\mathbf{e}_{z}-\gamma_{\rm S}\hat{\mathbf{K}})c/\sqrt{\varepsilon_{\rm S}}\omega$.

For the numerical integration, the integral over the angle $\mathbf{K}$ is performed analytically. The radial $K$ integral left in and the integrals over $\omega$ have to be performed numerically.

It is worth noting that $W_{ii}(\omega,T,\mathbf{r},\mathbf{r})$ is real and positive.

Then, it is straightforward to prove that:
\begin{equation}
g_{ii}(\omega,\mathbf{K})=\frac{i}{2\varepsilon_{0}\gamma}\left[\frac{\omega^{2}}{c^{2}}r_{s}+\left(K^{2}-\gamma^{2}\right)r_{p}\right]\equiv ig(\omega,K),
\end{equation}
\begin{equation}
\mathbb{W}_{ii}(\omega,\mathbf{K})=\frac{3\pi c}{4\omega}\frac{{\rm Re}(\gamma_{\rm S})}{|\gamma_{\rm S}|^{2}}\left[|t_{s}|^{2}+\frac{c^{4}}{|\varepsilon_{\rm S}|\omega^{4}}\left(K^{2}+|\gamma|^{2}\right)\left(K^{2}+|\gamma_{\rm S}|^{2}\right)|t_{p}|^{2}\right]\equiv\mathbb{W}(\omega,K),
\end{equation}
\noindent having that $\mathbb{W}(\omega,K)>0$ for every $\omega,K$.

And therefore, by deriving, changing to polar coordinates and integrating the angular variable:
\begin{equation}
\nabla\left[G_{ii}(\omega,\mathbf{r},\mathbf{r}')\right]|_{\mathbf{r}'=\mathbf{r}}=i\mathbf{e}_{z}\int_{0}^{+\infty}\frac{dK}{2\pi}~K\gamma~e^{i2\gamma z}ig(\omega,K),
\end{equation}
\begin{equation}
\nabla'\left[W_{ii}(\omega,\mathbf{r},\mathbf{r}')\right]|_{\mathbf{r}'=\mathbf{r}}=-\frac{i2\hbar\omega^{3}}{3\pi\varepsilon_{0}c^{3}}\mathbf{e}_{z}\int_{0}^{+\infty}\frac{dK}{2\pi}~K\gamma^{*}~e^{-2{\rm Im}(\gamma)z}\mathbb{W}(\omega,K).
\end{equation}

Finally, the force over the dipole results along the $z-$axis and only depending on $z$ (the distance of the nanoparticle to the plate), $\mathbf{F}\left(\mathbf{r}\right)=\mathbf{e}_{z}F_{z}(z)$ reads:
\begin{eqnarray}
F_{z}(z)&=&2\hbar\varepsilon_{0}\int_{0}^{+\infty}\frac{d\omega}{2\pi}\int_{0}^{+\infty}\frac{dK}{2\pi}K\Bigg[\coth\left[\frac{\hbar\omega}{2k_{B}T_{\rm EM}}\right]{\rm Re}\left[\alpha(\omega)\right]{\rm Re}\left[\gamma e^{i2\gamma z}ig(\omega,K)\right]\nonumber\\
&-&\coth\left[\frac{\hbar\omega}{2k_{B}T_{\rm NP}}\right]{\rm Im}\left[\alpha(\omega)\right]{\rm Im}\left[\gamma e^{i2\gamma z}ig(\omega,K)\right]\nonumber\\
&+&\left[\overline{n}_{\rm S}(\omega)-\overline{n}_{\rm EM}(\omega)\right]\frac{2\omega^{3}}{3\pi\varepsilon_{0}c^{3}}\left[{\rm Im}\left[\alpha(\omega)\right]{\rm Re}(\gamma)-{\rm Re}\left[\alpha(\omega)\right]{\rm Im}(\gamma)\right]e^{-2{\rm Im}(\gamma)z}\mathbb{W}(\omega,K)\Bigg].
\label{FzFinal}
\end{eqnarray}

At first sight, we can clearly differentiate the contributions due to the dispersive nature of the nanoparticle, which are all the terms including ${\rm Re}[\alpha(\omega)]$, and to the absorption of the nanoparticle, given by the terms including ${\rm Im}[\alpha(\omega)]$.

At the same time, we can eventually discriminate the contributions of the propagating and evanescent modes in vacuum. As $\gamma=\sqrt{\omega^{2}/c^{2}-K^{2}}$, the modes having $\omega,K$ such that $\gamma={\rm Re}(\gamma)$ are the propagating modes, while the modes such that $\gamma=i{\rm Im}(\gamma)$ are the evanescent modes. The second line in the last equation clearly presents the separation between them, where each set is correlated to the dispersion and absorption provided by the nanoparticle. In this term, the propagating modes are associated to the absorption of the nanoparticle while the evanescent ones are associated to the dispersion of the nanoparticle. On the other hadn, in the first line, the modes are not correlated with a specific property of the nanoparticle.

Another separation that is implicit in the form that Eq.(\ref{FzFinal}) is written relates to the Fresnel's coefficients associated to the plate. It is worth noting that, on one hand, $g$ only depends on the reflection coefficients $r_{\rm s,p}$, while $\mathbb{W}$ depends on the transmission coefficients $t_{\rm s,p}$. Thus, the first line of the r.h.s. of Eq.(\ref{FzFinal}) only depends on $r_{\rm s,p}$, while the second line on $t_{\rm s,p}$.

To simplify the expression, we can re cast it in the following way:
\begin{eqnarray}
F_{z}(z)&=&-\int_{0}^{+\infty}\frac{d\omega}{2\pi}\Bigg[\coth\left[\frac{\hbar\omega}{2k_{B}T_{\rm EM}}\right]{\rm Re}\left[\alpha(\omega)\right]{\rm Im}\left[\mathcal{R}(\omega,z)\right]+\coth\left[\frac{\hbar\omega}{2k_{B}T_{\rm NP}}\right]{\rm Im}\left[\alpha(\omega)\right]{\rm Re}\left[\mathcal{R}(\omega,z)\right]\nonumber\\
&+&\left[\overline{n}_{\rm EM}(\omega)-\overline{n}_{\rm S}(\omega)\right]\left[{\rm Im}\left[\alpha(\omega)\right]{\rm Re}[\mathcal{T}(\omega,z)]-{\rm Re}\left[\alpha(\omega)\right]{\rm Im}[\mathcal{T}(\omega,z)]\right]\Bigg],
\label{FzFinalReCast}
\end{eqnarray}
with:
\begin{equation}
\mathcal{R}(\omega,z)=2\hbar\varepsilon_{0}\int_{0}^{+\infty}\frac{dK}{2\pi}K\gamma~e^{i2\gamma z}~g(\omega,K)~~~,~~~\mathcal{T}(\omega,z)=\frac{4\hbar\omega^{3}}{3\pi c^{3}}\int_{0}^{+\infty}\frac{dK}{2\pi}K\gamma~e^{-2{\rm Im}(\gamma)z}~\mathbb{W}(\omega,K).
\end{equation}

Notice that written in this way, $\mathcal{R}$ only contains information on the reflection coefficients of the surface, while $\mathcal{T}$ depends on the transmission ones. We can also notice that at zero temperature, only the first line of Eq.(\ref{FzFinalReCast}) contributes to the force. Thus, we can split this zero-point fluctuations contribution to obtain Eq.(10) of the manuscript:
\begin{equation}
F_{z}(z)=F_{0}(z)+F_{\rm R}\left(z,T_{\rm EM},T_{\rm NP}\right)+F_{\rm T}\left(z,T_{\rm EM},T_{\rm S}\right),
\label{FzGeneral}
\end{equation}
with each contribution given by:
\begin{equation}
F_{0}(z)=-{\rm Im}\left[\int_{0}^{+\infty}\frac{d\omega}{2\pi}\alpha(\omega)\mathcal{R}(\omega,z)\right],
\label{FZeroPoint}
\end{equation}
\begin{equation}
F_{\rm R}\left(z,T_{\rm EM},T_{\rm NP}\right)=-2\int_{0}^{+\infty}\frac{d\omega}{2\pi}\left[\overline{n}_{\rm EM}(\omega){\rm Re}\left[\alpha(\omega)\right]{\rm Im}\left[\mathcal{R}(\omega,z)\right]+\overline{n}_{\rm NP}(\omega){\rm Im}\left[\alpha(\omega)\right]{\rm Re}\left[\mathcal{R}(\omega,z)\right]\right],
\label{FRTEMTNP}
\end{equation}
\begin{equation}
F_{\rm T}\left(z,T_{\rm EM},T_{\rm S}\right)=-\int_{0}^{+\infty}\frac{d\omega}{2\pi}\left[\overline{n}_{\rm EM}(\omega)-\overline{n}_{\rm S}(\omega)\right]\left[{\rm Im}\left[\alpha(\omega)\right]{\rm Re}[\mathcal{T}(\omega,z)]-{\rm Re}\left[\alpha(\omega)\right]{\rm Im}[\mathcal{T}(\omega,z)]\right].
\label{FTTEMTS}
\end{equation}

In general, the zero point contribution is obtained by means of Wick rotation on the complex $\omega-$plane under some analytical conditions of the integrand on the first quadrant. Provided that $\alpha(i\omega)\in\mathbb{R}$ and $\mathcal{R}(i\omega,z)\in\mathbb{R}$, it is common to obtain:
\begin{equation}
F_{0}(z)=-\int_{0}^{+\infty}\frac{d\omega}{2\pi}\alpha(i\omega)\mathcal{R}(i\omega,z).
\label{FZeroPointWick}
\end{equation}
This contribution depends only on the reflection coefficients of the surface, as it happens for the contribution $F_{\rm R}$, while $F_{\rm T}$ depends exclusively on the surface's transmission coefficients. Moreover, notice that by considering a scenario where the plane surface and the surrounding EM field are characterized by a unique environmental temperature ($T_{\rm EM} = T_{\rm S} \equiv T_{\rm Env}$) we have that $F_{\rm T}(z, T_{\rm Env}, T_{\rm Env}) = 0$, which shows that the force in this scenario only depends on the reflection coefficients but not on the transmission.

\section{The Casimir-Polder force for a metamaterial surface and the perfect (electric or magnetic) conductor limit}

The expression for the force on Eq.(\ref{FZeroPointWick}) is completely general for describing the EM interaction of a dipolar point-object placed at a distance $z$ from a material plate. As we mentioned before, the information about the nature of the plate is mainly encoded on the Fresnel's reflection and transmission coefficients $\{r_{\rm s,p};t_{\rm s,p}\}$. The surface's material in our work is given by Eq.(2) of the manuscript.

Before Wick rotation, for a full-bandwidth perfect conductor (electric or magnetic), the kernel $\mathcal{R}$ can be written as:
\begin{equation}
    \mathcal{R}(\omega,z)=\pm\mathcal{R}_{\rm PC}(\omega,z),
\end{equation}
with:
\begin{equation}
\mathcal{R}_{\rm PC}(\omega,z)= 2\hbar\int_{0}^{+\infty}\frac{dK}{2\pi}K e^{i2\gamma z}\gamma^{2}.
\end{equation}

Again, the two signs of the force are compatible with what it is intuitively expected from the interaction with image objects as described before.

It is straightforward that the total force for the PEC and PMC cases relate each other as:
\begin{equation}
    F_{z}^{\rm (PMC)}=-F_{z}^{\rm (PEC)},
\end{equation}
which also applies at zero-temperature and the short-distances regime.

At this point, Wick rotation ($\omega\rightarrow i\omega$) can be introduced for calculating the two contributions of kernel $\mathcal{R}$, obtaining:
\begin{equation}
    \mathcal{R}_{\rm PC}(i\omega,z)=2\hbar\int_{0}^{+\infty}\frac{dK}{2\pi}K e^{-2\sqrt{K^{2}+\frac{\omega^{2}}{c^{2}}} z}\left(K^{2}+\frac{\omega^{2}}{c^{2}}\right).
\end{equation}

By simple substitutions $K=(\omega/c)\sqrt{p^{2}-1}$, the integrals can be performed immediately:
\begin{equation}
    \mathcal{R}_{\rm PC}(i\omega,z)=\frac{\hbar}{\pi}\frac{\omega^{4}}{c^{4}}\int_{1}^{+\infty}dp~p^{3} e^{-2\frac{\omega}{c}p z}=-\frac{3\hbar}{8\pi z^{4}}A(i\omega,z)e^{-2\frac{\omega}{c}z},
    \label{RPC}
\end{equation}
with $A(i\omega,z)
=\sum_{n=0}^{3}\frac{1}{n!}\left(2\frac{\omega}{c}z\right)^{n}$.

After this simplification, we can see that the distinction on the contributions of the TE and TM modes has disappeared. This aspect is related to the fact that for a PC, all the modes are totally reflected without distinction.

%
%
%
%

Now, we turn to the frequency integration. For this, we start commenting on the properties of the polarizability. For a nanoparticle made of a material of permittivity $\varepsilon$, the polarizability is given by the Clausius-Mossoti formula:
\begin{equation}
\alpha(\omega)=V\xi(\omega)=3V\frac{[\varepsilon(\omega)-1]}{[\varepsilon(\omega)+2]},
\end{equation}
where $\xi$ corresponds to the polarizability per unit volume. In order to have a causal function $\alpha(t)$, the polarizability $\alpha(\omega)$ must have poles ($\{\nu_{m}\}$) lying on the lower-half of the complex plane (i.e., ${\rm Im}[\nu_{m}]<0~\forall~m$). Notice that we are not considering poles of $\alpha(\omega)$ on the real axis. Furthermore, for the typical material models, it happens that $\alpha(i\omega)\in\mathbb{R}~\forall~\omega$.

%
%

At the short-distance regime (defined by $k_{\rm B}Tz/[\hbar c]\ll 1$), the Casimir-Polder force between a point object and a plane surface is given by the zero-temperature contribution (see Ref.\cite{Buhmann1} and Sect.IIIA below), which according to Eq.(\ref{FZeroPointWick}) is given by:
\begin{equation}
    F_{0}(z)=\pm F_{0}^{\rm (PC)}(z).
    \label{ApproxFShort}
\end{equation}
where $\pm$ applies to the PMC and PEC cases respectively and having:
\begin{equation}
    F_{0}^{\rm (PC)}(z)=\frac{3\hbar V}{8\pi z^{4}}\int_{0}^{+\infty}\frac{d\omega}{2\pi}\xi(i\omega)A(i\omega,z)e^{-2\frac{\omega}{c}z},
\end{equation}

Notice that for on these two limiting cases we are not including the possibility of frequency cutoffs, for which the evaluation is numerical.

To go further, we consider a SiC nanoparticle of radius $R=50$nm in front of a full-bandwidth PMC surface. The permittivity model is $\varepsilon_{\rm SiC}(\omega)=\varepsilon_{\infty}(\omega_{\rm L}^{2}-\omega^{2}-i\gamma\omega)/(\omega_{\rm T}^{2}-\omega^{2}-i\gamma\omega)$, where $\omega_{\rm L} =18.253\times10^{13}{\rm s}^{-1}$, $\omega_{\rm T} =14.937\times10^{13}{\rm s}^{-1}$, $\gamma=8.966\times10^{11}{\rm s}^{-1}$ and $\varepsilon_{\infty}=6.7$. By a change of variables to the dimensionless variable $w=2\omega z/c$ we show that:
\begin{equation}
    F_{0}^{\rm (PC)}(z)=\frac{9\hbar c V}{32\pi^{2} z^{5}}\frac{(\varepsilon_{\infty}-1)}{(\varepsilon_{\infty}+2)}\mathcal{I}(z),
\end{equation}
with:
\begin{equation}
    \mathcal{I}(z)=\int_{0}^{+\infty}dw\frac{\left[w^{2}+\frac{2\gamma}{c}zw+\left(\frac{2z}{c}\right)^{2}\frac{(\varepsilon_{\infty}\omega_{\rm L}^{2}-\omega_{\rm T}^{2})}{(\varepsilon_{\infty}-1)}\right]}{\left[w^{2}+\frac{2\gamma}{c}zw+\left(\frac{2z}{c}\right)^{2}\frac{(\varepsilon_{\infty}\omega_{\rm L}^{2}+2\omega_{\rm T}^{2})}{(\varepsilon_{\infty}+2)}\right]}\left(1+w+\frac{w^{2}}{2}+\frac{w^{3}}{6}\right)e^{-w}.
\end{equation}

Now, it turns out that the integrand of $\mathcal{I}(z)$ decays approximately to 0 for $w>20$ and any $0<z<1\mu$m. Moreover, within the same range of distances $z$, we have that $\mathcal{I}(z)\approx\mathcal{I}(0)=4$ with a difference on the order of 3\%. Then, we have to a good level of approximation that:
\begin{equation}
    F_{0}^{\rm (PC)}(z)\approx \frac{9\hbar cV}{8\pi^{2}z^{5}}\frac{(\varepsilon_{\infty}-1)}{(\varepsilon_{\infty}+2)},
\end{equation}
which corresponds to the result shown in Eq.(8) of the manuscript.

Remarkably, the same expression is valid approximately (exactly) for Au (Si) by the replacement $\varepsilon_{\infty}\rightarrow\varepsilon_{\infty}^{\rm Au}(\varepsilon_{\rm Si})$.

\section{Features of the Casimir-Polder force for PMC surface}

\subsection{The non-equilibrium Casimir-Polder force and the short-distance regime}

In general scenarios for a nanoparticle in front of a plane surface, the non-equilibrium Casimir-Polder force is given by Eqs.(\ref{FzGeneral})-(\ref{FTTEMTS}). The expressions for the force can be re-cast in simple terms of Matsubara frequencies $\omega_{n}=2\pi nk_{B}T/\hbar$. For the thermal factors, the hyperbolic cotangent in terms of its poles located at $\omega_{n}$:
\begin{equation}
\coth\left(\frac{\hbar\omega}{2k_{B}T}\right)=\frac{2k_{B}T}{\hbar}\left[\frac{1}{\omega}+\sum_{n=1}^{+\infty}\left(\frac{1}{\omega+i\omega_{n}}+\frac{1}{\omega-i\omega_{n}}\right)\right].
\end{equation}

In addition, we restrict to the case of full-bandwidth perfect conductors for simplicity, so $F_{\rm T}\rightarrow 0$. We notice that $\mathcal{R}_{\rm PC}$ does not present poles in $\omega$. By employing a contour enclosing the first quadrant of the complex plane, we have for the integrals of interest:
\begin{eqnarray}
&&\int_{0}^{+\infty}\frac{d\omega}{2\pi}\coth\left(\frac{\hbar\omega}{2k_{B}T}\right){\rm Re}\left[\alpha(\omega)\right]\mathcal{R}_{\rm PC}(\omega,z)=\int_{0}^{+\infty}\frac{d\omega}{2\pi}\cot\left(\frac{\hbar\omega}{2k_{B}T}\right)\frac{\left[\alpha(i\omega)+\alpha^{*}(i\omega)\right]}{2}\mathcal{R}_{\rm PC}(i\omega,z)\nonumber\\
&&+~i\frac{k_{B}T}{2\hbar}\sum_{n=0}^{+\infty}\left(2-\delta_{n,0}\right)\frac{\left[\alpha(i\omega_{n})+\alpha^{*}(i\omega_{n})\right]}{2}\mathcal{R}_{\rm PC}\left(i\omega_{n},z\right)+\frac{i}{2}\sum_{\nu_{m}^{*}}R_{m}[\alpha^{*}]\coth\left(\frac{\hbar\nu_{m}^{*}}{2k_{B}T}\right)\mathcal{R}_{\rm PC}\left(\nu_{m}^{*},z\right),
\end{eqnarray}
\begin{eqnarray}
&&\int_{0}^{+\infty}\frac{d\omega}{2\pi}\coth\left(\frac{\hbar\omega}{2k_{B}T}\right){\rm Im}\left[\alpha(\omega)\right]\mathcal{R}_{\rm PC}(\omega,z)=\int_{0}^{+\infty}\frac{d\omega}{2\pi}\cot\left(\frac{\hbar\omega}{2k_{B}T}\right)\frac{\left[\alpha(i\omega)-\alpha^{*}(i\omega)\right]}{2i}\mathcal{R}_{\rm PC}(i\omega,z)\nonumber\\
&&+~\frac{k_{B}T}{2\hbar}\sum_{n=0}^{+\infty}\left(2-\delta_{n,0}\right)\frac{\left[\alpha(i\omega_{n})-\alpha^{*}(i\omega_{n})\right]}{2}\mathcal{R}_{\rm PC}\left(i\omega_{n},z\right)-\frac{1}{2}\sum_{\nu_{m}^{*}}R_{m}[\alpha^{*}]\coth\left(\frac{\hbar\nu_{m}^{*}}{2k_{B}T}\right)\mathcal{R}_{\rm PC}\left(\nu_{m}^{*},z\right),
\end{eqnarray}
with $R_{m}[\alpha^{*}]={\rm Res}[\alpha^{*}(\omega),\nu_{m}^{*}]$, with $\nu_{m}$ the poles of the polarizability $\alpha(\omega)$ as mentioned in the previous section. Notice that this calculations are valid for arbitrary scenarios (different temperatures). Employing the last expressions, we can write the Casimir-Polder force as for a PEC (PMC) surface:
\begin{equation}
F_{z}(z)=\pm\left[F_{\rm St}(z,T_{\rm EM})+F_{\rm Env}(z,T_{\rm EM})+F_{\rm Mat}(z,T_{\rm EM},T_{\rm NP})+F_{\rm Rad}(z,T_{\rm NP})\right],
\label{FNEqPC}
\end{equation}
with the different contributions given by:
\begin{equation}
F_{\rm St}(z,T_{\rm EM})=-\frac{3k_{B}T_{\rm EM}}{16\pi z^{4}}\alpha_{0},
\end{equation}
\begin{equation}
F_{\rm Env}(z,T_{\rm EM})=\frac{k_{B}T_{\rm EM}}{\hbar}\sum_{n=1}^{+\infty}\frac{\left[\alpha(i\omega_{n})+\alpha^{*}(i\omega_{n})\right]}{2}\mathcal{R}\left(i\omega_{n},z\right),
\end{equation}
\begin{equation}
F_{\rm Mat}(z,T_{\rm EM},T_{\rm NP})=\sum_{\nu_{m}^{*}}{\rm Re}\left[R_{m}[\alpha^{*}]\left(\mathcal{N}\left[\frac{\hbar\nu_{m}^{*}}{k_{B}T_{\rm EM}}\right]-\mathcal{N}\left[\frac{\hbar\nu_{m}^{*}}{k_{B}T_{\rm NP}}\right]\right)\mathcal{R}\left(\nu_{m}^{*},z\right)\right],
\end{equation}
\begin{equation}
F_{\rm Rad}(z,T_{\rm NP})=\frac{k_{B}T_{\rm NP}}{\hbar}\sum_{l=1}^{+\infty}\frac{\left[\alpha(i\omega_{l})-\alpha^{*}(i\omega_{l})\right]}{2}\mathcal{R}\left(i\omega_{l},z\right),
\end{equation}
where $\alpha_{0}\equiv\alpha(0)$ stands for the static polarizability, the Matsubara $\omega_{n}=2\pi nk_{B}T_{\rm EM}/\hbar$ while $\omega_{l}=2\pi lk_{B}T_{\rm NP}/\hbar$, and with $\mathcal{N}(x)=1/({\rm Exp}[x]-1)$.

First, the contribution corresponding to $F_{\rm St}$ stands for the thermal Casimir-Polder force between the surface and the nanoparticle. This contribution is independent of the material of the plate and depends only on the static value of the polarizability ($\alpha_{0}$), while it depends on the temperature of the environmental EM field $T_{\rm EM}$. Secondly, $F_{\rm Env}$ corresponds to the contribution of the environmental EM field including the dynamical properties of the nanoparticle and also the characteristics of the plate. This contribution is responsible of the Casimir-Polder force at thermal equilibrium at zero temperature and the fact that depends on the combination $\alpha+\alpha^{*}$ resembles the fact that depends on the real part of the polarizability which is associated to the dispersive properties of the nanoparticle. This is consistent with the fact that this contribution depends on the temperature of the environmental EM field, which is dispersed by the nanoparticle. The contribution $F_{\rm Mat}$ stands for a purely non-equilibrium contribution since at equilibrium $F_{\rm Mat}(z,T,T)=0$ for arbitrary $T$. Including the information of the poles of the polarizability, this contribution is characterized by the plasmonic frequencies ${\rm Re}(\nu_{m})$ of the material nanoparticle. Lastly, the contribution in $F_{\rm Rad}$ stands for a force associated to the radiation of the nanoparticle. The fact that depends on the combination $\alpha-\alpha^{*}$ resembles the fact that depends on the imaginary part of the polarizability which is associated to the dissipative (absorption and emission) properties of the nanoparticle. This is consistent with the fact that this contribution depends on the temperature of the nanoparticle.

Finally, notice that in thermal equilibrium ($T_{\rm EM}=T_{\rm NP}\equiv T$), the expression simplifies to:
\begin{eqnarray}
F_{z}^{\rm Eq}(z)&=&\pm\left[F_{\rm St}(z,T)+F_{\rm Env}(z,T)+F_{\rm Rad}(z,T)\right]\nonumber\\
&=&\pm\left[-\frac{3k_{B}T_{\rm EM}}{16\pi z^{4}}\alpha_{0}+\frac{k_{B}T_{\rm EM}}{\hbar}\sum_{n=1}^{+\infty}\alpha(i\omega_{n})\mathcal{R}_{\rm PC}\left(i\omega_{n},z\right)\right],
\label{FEqPC}
\end{eqnarray}
which for the PEC case stands for the typical Casimir-Polder force at thermal equilibrium.

\begin{figure}[!ht]
\includegraphics[width=0.7\columnwidth]{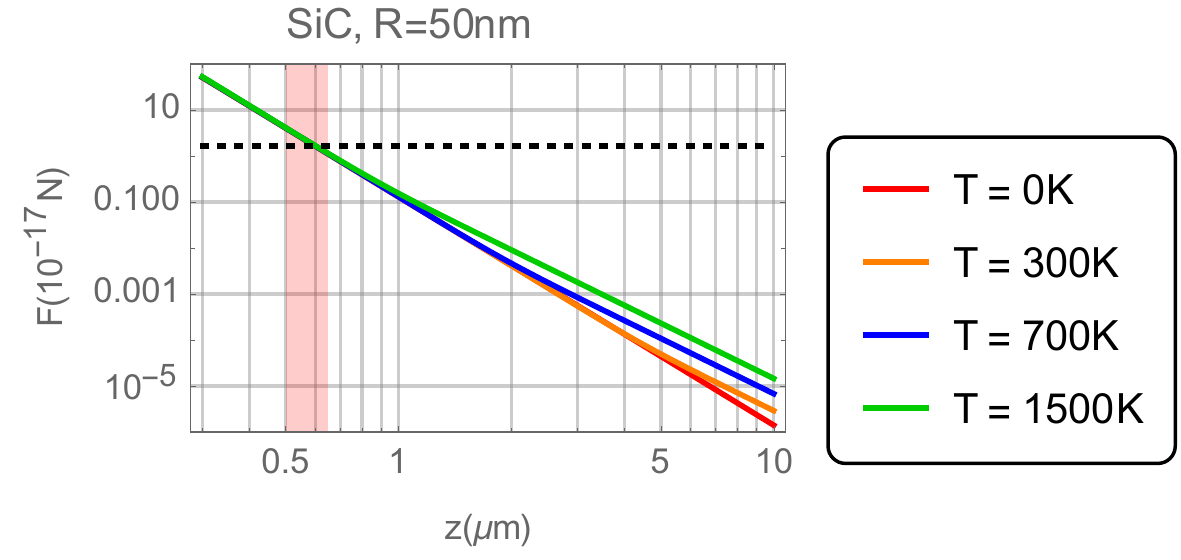}
\caption{\label{fig:ContributionZeroT} Casimir-Polder for acting on a SiC nanoparticle with $R=50$nm radius in front of a PMC plane surface and held at equilibrium for temperature values $T=0{\rm K},300{\rm K},700{\rm K},1500{\rm K}$, according to Eq.(\ref{ApproxFShort}) (with $b=0$ and no cutoffs) and Eq.(\ref{FEqPC}). The black dotted horizontal line corresponds to the modulus of the nanoparticle's weight force $mg$. The intersections between the weight and the different curves give the levitation position for each case. The red shaded region corresponds to an approximate region 0.5-0.65$\mu$m corresponding to the levitation of SiC nanoparticles in front of broadband PMCs.}
\end{figure}

For a SiC nanoparticle of radius $R=50$nm in front of a PMC surface, we show in Fig.\ref{fig:ContributionZeroT} the equilibrium Casimir-Polder force for different temperatures.
We can clearly observe that for $z<1\mu$m (or in the short-distance regime defined by $k_{\rm B}Tz/[\hbar c]\ll 1$) all the curves are almost the same for a broad range of temperature values. In this sense, it is verified that in the short-distance regime we have $F_{z}(z)\approx F_{0}(z)$ as mentioned in the manuscript.

\subsection{Force on SiC, Au and Si nanoparticles}

\begin{figure}[!ht]
\includegraphics[width=0.45\columnwidth]{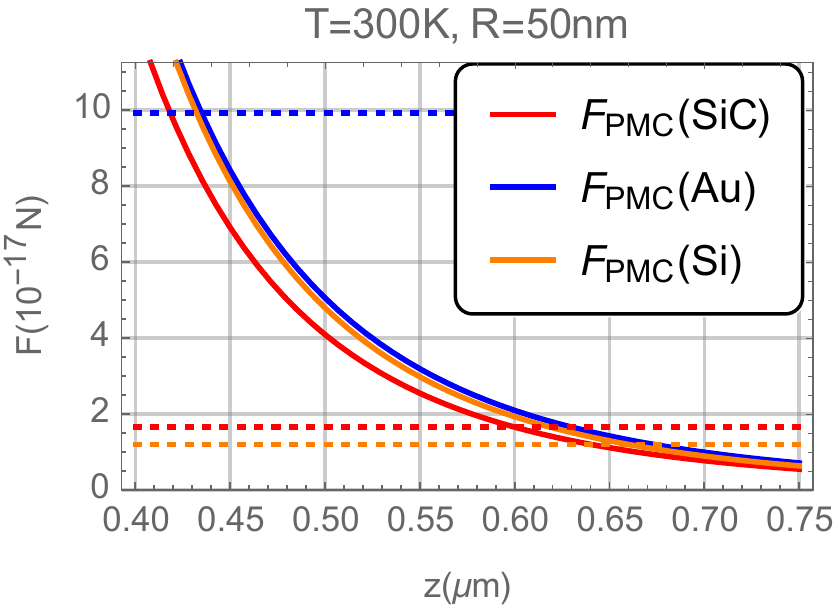}
\caption{\label{fig:Features}Equilibrium Casimir-Polder forces at $T=300$K for SiC (red lines), Au (blue lines) and Si (orange lines) nanoparticles in front of PEC and PMC surfaces. The curves correspond to Eq.(\ref{FEqPC}), which at the short-distances regime also corresponds to Eq.(\ref{ApproxFShort}) with $b=0$. Solid curves correspond to the interaction with PMC surfaces, while dashed curves correspond to PEC surfaces. Dotted horizontal lines corresponds to the weights of each nanoparticle.}
\end{figure}

The Casimir-Polder force for the PMC plane depends on the nanoparticle's polarizability according to Eq.(\ref{FEqPC}). We consider the levitation of SiC, Au and Si nanoparticles. For SiC we employed a permittivity model $\varepsilon_{\rm SiC}(\omega)=\varepsilon_{\infty}(\omega_{\rm L}^{2}-\omega^{2}-i\gamma\omega)/(\omega_{\rm T}^{2}-\omega^{2}-i\gamma\omega)$, where $\omega_{\rm L} =18.253\times10^{13}{\rm s}^{-1}$, $\omega_{\rm T} =14.937\times10^{13}{\rm s}^{-1}$, $\gamma=8.966\times10^{11}{\rm s}^{-1}$ and $\varepsilon_{\infty}=6.7$. For Au we take $\varepsilon_{\rm Au}(\omega)=\varepsilon_{\infty}^{\rm Au}-\omega_{\rm Pl}^{2}/(\omega^{2}+i\gamma_{\rm Au}\omega)$, with $\omega_{\rm Pl}=2.15\times10^{15}{\rm s}^{-1}$, $\gamma_{\rm Au}=5.88\times10^{13}{\rm s}^{-1}$ and $\varepsilon_{\infty}^{\rm Au}=5$. For Si we used a constant permittivity $\varepsilon_{\rm Si}=12.25$ (corresponding to a refraction index of 3.5). The mass densities for each material are $\rho_{\rm SiC}=3210K{\rm g}/{\rm m}^{3}$, $\rho_{\rm Au}=19300K{\rm g}/{\rm m}^{3}$, $\rho_{\rm Si}=2330K{\rm g}/{\rm m}^{3}$. In Fig.\ref{fig:Features} we show the comparison between the Casimir-Polder forces and the weights corresponding to these types of material nanoparticles.
The intersections of a colored curve with the respective dotted horizontal lines corresponds to the equilibrium points. For SiC the equilibrium occurs at $z_{\rm SiC}=600$nm, for Au we have $z_{\rm Au}=440$nm and for Si $z_{\rm Si}=660$nm.